\documentstyle[12pt]{article}
\raggedright 
\renewcommand{\section}[1]{\refstepcounter{section}
\vspace{24pt}\noindent{\bf\arabic{section}.\quad #1}
\vspace*{12pt}}
\newcommand{\ulsect}[1]{\vspace{18pt}\noindent{\bf #1}
\vspace*{12pt}}

\begin{document}

\begin{flushright} CERN-TH.6892/93\\
\end{flushright}
\vspace*{10mm}
\begin{center}
{\bf Undesirable effects of covariance matrix techniques for error
analysis}\\[10mm]
David Seibert$^*$\\[5mm] Theory Division, CERN, CH-1211 Geneva 23,
Switzerland\\[10mm]
{\bf Abstract}\\
\end{center}
\hspace*{12pt}Regression with $\chi^2$ constructed from the covariance
matrix should not be used for some combinations of covariance matrices
and fitting functions.  Using the technique for unsuitable combinations
can amplify systematic errors.  This amplification is uncontrolled, and
can produce arbitrarily inaccurate results that might not be ruled out
by a $\chi^2$ test.  In addition, this technique can give incorrect
(artificially small) errors for fit parameters.  I give a test for this
instability and a more robust (but computationally more intensive)
method for fitting correlated data.\\
\vfill
\begin{center}
{\em Submitted to Phys.\ Rev.\ D}
\end{center}
\vfill
CERN-TH.6892/93\\
May 1993\\
(Revised September 1993)\\
\vspace*{10mm}
\footnoterule
\vspace*{3pt}
$^*$On leave until October 12, 1993 from:
Physics Department, Kent State University, Kent, OH 44242 USA.
Internet: seibert@surya11.cern.ch.\\

\newpage\setcounter{page}{1} \pagestyle{plain}
     \setlength{\parindent}{12pt}

Recently there has been some interest in the analysis of correlated
data, and people seeking more sophisticated analysis techniques have
often performed regression, using the covariance matrix to construct
$\chi^2$ [1].  DeGrand [2], DeTar and Kogut [3] and Gottlieb
{\it et al}. [4] use the technique to analyze lattice gauge theory
results, while Abreu {\it et al}.\ [5] and Wosiek [6] use it to
analyze scaled factorial moment data.  I show here that this analysis
technique can amplify systematic errors, unlike simpler, more robust
techniques.

This technique is simple in principle -- transform the data to an
uncorrelated basis, use regression to fit the data in this basis, then
transform back to the laboratory frame.  However, some of the results
obtained by this procedure are very odd.  In particular, Gottlieb
{\it et al}. [4], Toussaint [7] and Wosiek [6] find that this procedure
can produce best-fit lines that fall below all data points, and even
below all error bars!

In this paper, I first discuss the proposed treatment of correlated
data, and show that in a {\em gedanken} experiment without systematic
errors this treatment produces exactly the desired results.  In a very
similar {\em gedanken} experiment with arbitrarily small systematic
errors, this procedure amplifies the errors in the data; therefore,
this treatment of data is not robust.  I use these simple {\em gedanken}
experiments instead of the scaled factorial moment data or the lattice gauge
theory results for purposes of presentation, as the effect observed in the
different data sets is qualitatively the same.  I then give a more robust
alternative procedure for fitting correlated data, and in the course of
this discussion a test for the stability of the regression is shown.

The experimental procedure is very simple.  Consider $N$ trial
measurements of $I$ data points, $y_i$.  Calculate the covariance matrix
from these data,
\begin{equation}
C_{ij} = \frac {1} {N-1} \left[ \sum_{n=1}^N (y_{i,n} -y_i) (y_{j,n} -y_j)
\right]~, \label{eCij}
\end{equation}
where
\begin{equation}
y_i ~=~ \frac {1} {N} \sum_{n=1}^N y_{i,n}~.
\end{equation}
Fit the data with the curve $f_i ( \alpha )$, where $\{ \alpha \}$ is the
set of free parameters, by minimizing
\begin{equation}
\chi^2 = \sum_{i=1}^I \sum_{j=1}^I \, (y_i -f_i) \,
(C^{-1})_{ij} (y_j -f_j)~. \label{echi2}
\end{equation}

I illustrate this procedure with a {\em gedanken} experiment to measure
the mean voltage of a generator that produces random voltages $v$ with
probability distribution $p(v)$.  The generator charges a capacitor, and
I then measure the voltage on the capacitor twice, calling the two
measurements $y_1$ and $y_2$.  Each measurement has some (uncorrelated)
``measurement noise'' in addition to the fluctuations due to the random
voltage generator; I assume that this noise may be different for the two
measurements.

After $N$ trials, the experimentally determined covariance matrix is
\begin{equation}
C = \frac {1} {N-1} \left( \begin{array} {cc}
                           \sigma^2 + e_1^2   & \sigma^2 \\
                           \sigma^2           & \sigma^2 + e_2^2
                \end{array} \right)~, \label{eCija}
\end{equation}
where
\begin{equation}
\sigma^2 = \int dv \, p(v) \, v^2 - \left[ \int dv \, p(v) \, v
\right]^2 \label{esp}
\end{equation}
is the contribution to the covariance matrix from the distribution of
random voltages, and $e_{1(2)}^2$ is the contribution of noise from
the set of first (second) measurements.  Fitting to the function $f_1 =
f_2 = V$, $\chi^2$ is minimized when
\begin{equation}
V = \frac {e_1^{-2} \langle y_1 \rangle + e_2^{-2} \langle
y_2 \rangle} {e_1^{-2} + e_2^{-2}}~, \label{eV1}
\end{equation}
where $\langle y_{1(2)} \rangle$ is the average value for the first
(second) measurement.  It is clear from Eq.~(\ref{eV1}) that $V$ is the
average of $\langle y_1 \rangle$ and $\langle y_2 \rangle$, properly
weighted for measurement error, and so the analysis procedure is very
successful at fitting the curve for this {\em gedanken} experiment.

The experimental error is given by
\begin{equation}
\sigma_V^2 = 2 \left( \frac {d^2} {dV^2} \chi^2 \right)^{-1}.
\end{equation}
Taking $\chi^2$ from eq.~(\ref{echi2}) and $C$ from eq.~(\ref{eCija})
gives
\begin{equation}
\sigma_V^2 = \frac {\sigma^2 + 1 / (e_1^{-2}+e_2^{-2})} {N-1}.
\end{equation}
Again, this technique works well, clearly giving the correct error in
the cases $\sigma = 0$ and $\sigma \rightarrow \infty$.

Now, I modify the {\em gedanken} experiment slightly, by assuming that
the capacitor discharges somewhat between the two measurements.  I
therefore assume that the first measurement is unchanged, but that the
voltage is reduced by a factor $\gamma$ at the time of the second
measurement.  I could alternatively assume that the scales of the
voltmeters are slightly different, but I wish to have all systematic
effects occur before measurement rather than during it.

In this case, the experimentally determined covariance matrix is
\begin{equation}
C = \frac {1} {N-1} \left( \begin{array} {cc}
                  \sigma^2 + e_1^2   & \gamma \sigma^2 \\
                  \gamma \sigma^2    & \gamma^2 \sigma^2 + e_2^2
                \end{array} \right)~. \label{eCijb}
\end{equation}
After an infinite number of measurements $\langle y_1 \rangle = \overline{v}$
and $\langle y_2 \rangle = \gamma \overline{v}$, where $\overline{v}$ is the
mean value of the random voltage.  Fitting again to the function $f_1 = f_2 =
V$, $\chi^2$ is minimized when
\begin{equation}
V = \frac {\gamma e_1^2 + e_2^2} {(\gamma-1)^2 \sigma^2 +e_1^2 +e_2^2} \,
\overline{v}~. \label{eVb}
\end{equation}

The procedure gives {\em systematic} errors for this second experiment,
as $V$ is always less than $\overline{v}$.  This is not totally
unexpected, because the discharge of the capacitor between the
measurements gives a systematically lower value of $y_2$.  If
$\gamma=1$, as in the first experiment, all systematic errors vanish.
For any other value of $\gamma$, however, $V$ can have any value
between zero and $\overline{v}$.  By contrast, a naive least-squares fit
always yields a value for $V$ between $\gamma \overline{v}$ and
$\overline{v}$.  Thus, the covariance matrix technique can produce large
systematic errors from arbitrarily small intrinsic systematic errors.

One might think that $\chi^2$ should be large whenever the fit is very
bad ($V \ll \overline{v}$).  However, this is not the case if the sample
size is too small.  In the limit $\sigma \rightarrow \infty$, where the
fit is the worst,
\begin{equation}
\chi^2 \rightarrow (N-1) \left( \frac {\overline{v}} {\sigma} \right)^2.
\end{equation}
Thus, $\chi^2$ will be acceptably small whenever
$N < (\sigma/\overline{v})^2$, so that an infinite number of events
may be required to rule out the worst fits.

One might then expect that, if $\chi^2$ is acceptably small, the error
in $V$ will be large enough that $V$ is within a few standard
deviations of $\overline{v}$.  However, in the limit $(\gamma-1)\sigma
\gg e_1,e_2$,
\begin{equation}
\sigma_V^2 = \frac {(\gamma^2 e_1^2+e_2^2) \sigma^2 + e_1^2e_2^2}
{(N-1) \left[(\gamma-1)^2 \sigma^2 +e_1^2 +e_2^2 \right]}
\rightarrow \frac {\gamma^2 e_1^2+e_2^2} {(N-1) (\gamma-1)^2}
\ll \frac {\sigma^2} {N-1}.
\end{equation}
Thus, it is quite possible to have simultaneously $V \ll \overline{v}$,
$\chi^2$ small, and $(V-\overline{v})^2 \gg \sigma_V^2$.

Now I try a more robust technique, constructing the best estimator by
minimizing the variance in
\begin{equation}
V = a y_1 + (1-a) y_2.
\end{equation}
The variance is
\begin{eqnarray}
\sigma_V^2 &=& \langle V^2 \rangle - \langle V \rangle^2, \\
&=& a^2 \left( \langle y_1^2 \rangle - \langle y_1 \rangle^2 \right)
+2a(1-a) \left( \langle y_1 y_2 \rangle - \langle y_1 \rangle
\langle y_2 \rangle \right)
+(1-a)^2 \left( \langle y_2^2 \rangle - \langle y_2 \rangle^2 \right), \\
&=& \frac {1} {N-1} \left\{ a^2 \left[ (\gamma-1)^2 \sigma^2 +e_1^2 +e_2^2
\right] -2a \left[ \gamma(\gamma-1) \sigma^2 +e_2^2 \right]
+\left[ \gamma^2 \sigma^2 + e_2^2 \right] \right\}.
\end{eqnarray}
The condition $d\sigma_V^2/da=0$ then gives
\begin{eqnarray}
a &=& \frac {\gamma(\gamma-1) \sigma^2 +e_2^2}
{(\gamma-1)^2 \sigma^2 +e_1^2 +e_2^2}, \\
V &=& \frac {\gamma e_1^2 +e_2^2} {(\gamma-1)^2 \sigma^2 +e_1^2 +e_2^2}
\overline{v}, \\
\sigma_V^2 &=& \frac {\left( \gamma e_1^2 + e_2^2 \right)^2 \sigma^2
+\left( \gamma(\gamma-1) \sigma^2 +e_2^2 \right)^2 e_1^2
+\left( (\gamma-1) \sigma^2 -e_1^2 \right)^2 e_2^2}
{(N-1) \left[ (\gamma-1)^2 \sigma^2 +e_1^2 +e_2^2 \right]^2}.
\end{eqnarray}

The value of $V$ is the same for the two techniques, and $\sigma_V^2$ is
the same when $\gamma=1$.  In the limit $e_1,\ e_2 \rightarrow 0$ I find
\begin{equation}
\sigma_V^2 = \frac {\gamma^2 e_1^2 +e_2^2} {(N-1) (\gamma-1)^2},
\label{eVbe0} \end{equation}
which is identical to the result obtained from regression.  Thus, the
techniques are almost the same.  However, the best estimator technique is
more transparent, and the cause of the instability is more easily
recognized and corrected with this technique.

In the previous analysis I left out a condition --- $a$ and $(1-a)$ must
both be non-negative.  In this case, the solution (\ref{eVbe0}) is only
valid when
\begin{eqnarray}
e_1^2 &\geq& (\gamma-1) \sigma^2, \\
e_2^2 &\geq& -\gamma(\gamma-1) \sigma^2.
\end{eqnarray}
Applying this condition, $\sigma_V^2$ is minimized with
\begin{equation}
a= \left\{ \begin{array} {ll}
     0 \quad & \gamma < 1, \\
     1 \quad & \gamma > 1,
   \end{array} \right.
\end{equation}
in the limit $e_1,\ e_2 \rightarrow 0$.  I then obtain
\begin{equation}
V = \left\{ \begin{array} {ll}
     \gamma \overline{v} \quad & \gamma < 1, \\
     \overline{v} \quad & \gamma > 1,
   \end{array} \right.
\end{equation}
and
\begin{equation}
\sigma_V^2 = \left\{ \begin{array} {ll}
     \gamma^2\sigma^2 +e_2^2 \quad & \gamma < 1, \\
     \sigma^2 +e_1^2 \quad & \gamma > 1.
   \end{array} \right.
\end{equation}
Thus, the systematic error is not amplified with this procedure, and
the estimate of $\sigma_V^2$ is not artificially small.

The crucial point is the non-negativity of $a$ and $1-a$.
Mathematically, this can be written as
\begin{equation}
\forall{}i:\ \frac {\partial f_i} {\partial y_i} \geq 0.
\end{equation}
This general requirement for a stable fit is that, given a perturbation
in the data, the function does not move locally against the direction of
the perturbation.  It is intuitively obvious, though I am not sure
whether it has been rigorously demonstrated.

The partial derivative is calculated as follows.  The general fitting
condition of minimizing $\chi^2$ can be written as
\begin{equation}
\forall{}a:\ \sum_{j,k} \left( C^{-1} \right)_{jk} \frac {\partial f_j}
{\partial \alpha_a} \left( f_k-y_k \right) = 0,
\end{equation}
where $\{ \alpha \}$ is the set of fitting parameters.  If $y_i \rightarrow
y_i+\delta y_i$, we must have now
\begin{equation}
\forall{}a:\ \sum_{j,k,b} \left( C^{-1} \right)_{jk} \left\{
\frac {\partial f_j} {\partial \alpha_a} \frac {\partial f_k} {\partial
\alpha_b} + \frac {\partial^2 f_j} {\partial \alpha_a \partial \alpha_b}
(f_k-y_k) \right\} \delta \alpha_b - \sum_j \left( C^{-1} \right)_{ij}
\frac {\partial f_j} {\partial \alpha_a} \delta y_i = 0.
\end{equation}
This can be written more compactly in matrix form:
\begin{eqnarray}
\delta \alpha_b &=& \sum_a \left( M^{-1} \right)_{ab} K_{ai} \delta y_i, \\
M_{ab} &=& \sum_{j,k} \left( C^{-1} \right)_{jk} \left\{
\frac {\partial f_j} {\partial \alpha_a} \frac {\partial f_k} {\partial
\alpha_b} + \frac {\partial^2 f_j} {\partial \alpha_a \partial \alpha_b}
(f_k-y_k) \right\}, \\
K_{ai} &=& \sum_j \left( C^{-1} \right)_{ij}
\frac {\partial f_j} {\partial \alpha_a}.
\end{eqnarray}
Finally, I obtain
\begin{eqnarray}
\delta f_i &=& \sum_b \frac {\partial f_i} {\partial \alpha_b}
\delta \alpha_b, \\
&=& \sum_{a,b} \frac {\partial f_i} {\partial \alpha_b}
\left( M^{-1} \right)_{ab} K_{ai} \delta y_i,
\end{eqnarray}
and the partial derivative is
\begin{equation}
\frac {\partial f_i} {\partial y_i} = \sum_{a,b} \frac {\partial f_i}
{\partial \alpha_b} \left( M^{-1} \right)_{ab} K_{ai}.
\end{equation}

For the fit to a constant $V$, $\partial f_i / \partial V = 1$, so
\begin{equation}
\frac {\partial f_i} {\partial y_i} = \frac
{\sum_j \left( C^{-1} \right)_{ij}} {\sum_{j,k} \left( C^{-1} \right)_{jk}}.
\end{equation}
The denominator is never negative, as it is equal to a sum of eigenvalues
of $C^{-1}$ (with all weights non-negative), and all eigenvalues of $C^{-1}$
are non-negative.  Thus, the stability condition for this regression is
\begin{equation}
\forall{}i:\ \sum_j \left( C^{-1} \right)_{ij} \geq 0,
\end{equation}
which is trivially satisfied for uncorrelated data ($C$ is then diagonal).
If this condition is violated, then the best estimator should be used
instead of the regression, to obtain the variance in the fit parameters.

The best estimator technique can also be used to fit lines and more
complicated curves to data.  For a line, first fit $y=ax+b$ to all
independent sets of points $ij$ to obtain
\begin{eqnarray}
a_{ij} &=& \frac {y_i-y_j} {x_i-x_j}, \\
b_{ij} &=& \frac {x_iy_j - x_jy_i} {x_i-x_j}.
\end{eqnarray}
Then construct linear estimators for the quantities $a$ and $b$,
\begin{eqnarray}
a &=& \sum_{i \neq j} k_{ij} a_{ij}, \\
b &=& \sum_{i \neq j} l_{ij} b_{ij},
\end{eqnarray}
with the constraints
\begin{equation}
\sum_{i \neq j} k_{ij} = \sum_{i \neq j} l_{ij} = 1.
\end{equation}
Finally, minimize the variance in $a$ and $b$, to obtain the values and
variances of both, but with the conditions
\begin{equation}
\forall{}ij:\ k_{ij}, l_{ij} >0.
\end{equation}
In general the procedure is not worth the effort required, as the fit
parameters are identical to those obtained with regression.

I have shown that covariance matrix regression should be supplemented
by a test for the stability of the regression.  When the regression is
unstable, the fit parameters can be altered in an uncontrolled fashion.
These alterations can sometimes be ruled out by a $\chi^2$ test; however,
for arbitrarily small $\chi^2$, if the data set is small and fluctuations
are large, the apparent errors in fit parameters can be much smaller than
the difference between their apparent values and the best estimators for
these values.

The alternative to using covariance matrix regression is to fit all
possible sets of points (as many points per set as there are fit
parameters) to obtain all possible linearly independent sets of the fit
parameters,and use the linear combinations of the values obtained in this
way (with no negative multipliers) that have the lowest variances as the
best estimators of the fit parameters.  This is computationally more
cumbersome, but is the more rigorous procedure so it may be simplest to
use this in the first place rather than attempting covariance matrix
regression first.

\bigskip

I thank F. James for helpful suggestions and K. Zalewski for useful
discussions.  This material is based upon work supported by the North
Atlantic Treaty Organization under a Grant awarded in 1991.


\ulsect{References}

\begin{list}{\arabic{enumi}.\hfill}{\setlength{\topsep}{0pt}
\setlength{\partopsep}{0pt} \setlength{\itemsep}{0pt}
\setlength{\parsep}{0pt} \setlength{\leftmargin}{\labelwidth}
\setlength{\rightmargin}{0pt} \setlength{\listparindent}{0pt}
\setlength{\itemindent}{0pt} \setlength{\labelsep}{0pt}
\usecounter{enumi}}

\item W.T. Eadie {\it et al}., {\it Statistical Methods in Experimental
Physics} (North-Holland, Amsterdam, 1971), pp.\ 62--66.

\item T.A. DeGrand, Phys.\ Rev.\ D {\bf 36}, 176 (1987).

\item C. DeTar and J.B. Kogut, Phys.\ Rev.\ D {\bf 36}, 2828 (1987).

\item S. Gottlieb {\it et al}., Phys.\ Rev.\ D {\bf 38}, 2245 (1988).

\item Abreu, P. {\it et al}., Phys.\ Lett.\ B {\bf 247}, 137 (1990).

\item B. Wosiek, Acta Physica Polonica {\bf B21}, 1021 (1990).

\item D. Toussaint, in {\it From Actions to Answers -- Proceedings of
the 1989 Theoretical Advanced Study Institute in Elementary Particle
Physics} (World Scientific, Singapore, 1990; T. DeGrand and D.
Toussaint, eds.).

\end{list}

\vfill \eject

\end{document}